\documentclass[twocolumn,pre,superscriptaddress,epfs,aps,showpacs]{revtex4}
\usepackage{amsmath}
\usepackage{graphicx}
\usepackage{color}
\usepackage{epsfig}
\usepackage{braket}	
\usepackage{latexsym}
\usepackage{array} 
\usepackage{multirow}

\begin{document}

\title{Controlling the transport of active matter in disordered lattices of asymmetrical obstacles}
\author{A. D. Borba}
%\email{andreduarte@fisica.ufc.br}
\affiliation{Departamento de F\'isica, Universidade Federal do
Cear\'a, Caixa Postal 6030, Campus do Pici, 60455-760 Fortaleza,
Cear\'a, Brazil}
\author{Jorge L. C. Domingos}
\email{jorgecapuan@fisica.ufc.br}
%\email{wandemberg@fisica.ufc.br}
\affiliation{Departamento de F\'isica, Universidade Federal do
Cear\'a, Caixa Postal 6030, Campus do Pici, 60455-760 Fortaleza,
Cear\'a, Brazil}
\author{E. C. B. Moraes}
\affiliation{Instituto Federal de Educa\c{c}\~ao, Ci\^encia e Tecnologia, Coordena\c{c}\~ao de Ensino M\'edio, Tucuru{\'\i}, Par\'a, Brasil}
\affiliation{Universidade Federal do Par\'a, Faculdade de F\'\i sica, ICEN, Av. Augusto Correa, 1, Guam\'a, 66075-110, Bel\'em, Par\'a, Brazil}
\author{F. Q. Potiguar}
\affiliation{Universidade Federal do Par\'a, Faculdade de F\'\i sica, ICEN, Av. Augusto Correa, 1, Guam\'a, 66075-110, Bel\'em, Par\'a, Brazil}
\author{W. P. Ferreira}
%\email{wandemberg@fisica.ufc.br}
\affiliation{Departamento de F\'isica, Universidade Federal do
Cear\'a, Caixa Postal 6030, Campus do Pici, 60455-760 Fortaleza,
Cear\'a, Brazil}

\begin{abstract}
We investigate the transport of active matter in the presence of a disordered square lattice of asymmetric obstacles, which is built by removing a fraction of them from the initial full lattice. We consider no external field. We observe a spontaneous inversion of the net particle current, compared to the usual sense of such a current reported in the literature, if the obstacle (half-circle) has the same diameter of the unit cell of the square lattice. If this diameter is smaller, there is no inversion. We show a calculation that reproduces our numerical results, based on the argument that such effects are a consequence of the imbalance of particles traveling in the positive and the negative directions due to traps formed by the obstacles: for positive travelers the traps are the spaces between neighboring obstacles, while for negative travelers, they are the flat side of the obstacles.
\end{abstract}


\maketitle
\section{Introduction}
{\color{black}Active Matter is the generic definition for particles that generate motion either by consuming their internal energy or even by utilizing the energy from the environment}\cite{toner05,ramaswamy10,vicsek12,marchetti12,bechinger16}. Physical models that simulate such particles are divided between flocking (Vicsek model) \cite{vicsek95,drocco12-02}, and angular Brownian motion (ABM) types \cite{fily12} (which also includes run-and-tumble dynamics - RTD \cite{wan08,tailleur09}). Among the various phenomena associated with such systems, the transport properties are the most investigated ones. It is known to be possible to rectify these particles motion when they are in an environment in which there is an intrinsic asymmetry, such as a regular lattice of funnel-shaped \cite{galajda07,wan08,drocco12-02,ghosh13,reichhardt13} or half-circular obstacles \cite{potiguar14}.

{\color{black}Active systems exhibit rich and intriguing nonequilibrium properties, including emergent structures with collective behavior distinguishing from that of the individual constituents.} More recently, there has been an increasing interest in the collective behavior of active matter in a disordered environment, defined as a system in which translational invariance is broken in some way. Some examples are seen in \cite{chepizhko13-1,dolai2018phase,chepizhko13-2,reichhardt14,morin16,wang17,yllanes2017}. Here, the invariance is broken by adding several fixed, randomly positioned, obstacles (in contrast, \cite{chepizhko13-1,dolai2018phase} also deals with moving obstacles). {\color{black}Other} interesting features are observed, {\color{black}ranging} from the existence of an optimal angular noise value that maximizes particle motion {\color{black} as a function  of the obstacle density} \cite{chepizhko13-1}, hindering of the particle's motion (trapping \cite{chepizhko13-2}, clogging \cite{reichhardt14}, and flocking suppression \cite{morin16}) to a rich relation between the mean search time of a target with the disorder \cite{wang17}. We propose to tackle a similar problem, the collective behavior of active particles in a disordered environment. We start from an {\em ordered} lattice of obstacles and we introduce the disorder by randomly removing a fraction of them. Our main interest is to investigate the transport of particles in a set up in which the translational invariance is broken both locally, due to the asymmetric obstacles, and globally, due to the resulting random arrangement of obstacles. In \cite{potiguar14}, it was shown that particles travel along the curvature of the half-circles, i.e., they follow the direction of the normal to their flat side (from now on, the easy flow direction) when the half-circle obstacles are arranged in a regular lattice. Our main observation is that, for low disorder, there is a {\em spontaneous inversion} of the current, i.e., particles tend to move in the negative easy flow direction with no external fields, which is opposite, for instance, to what was obtained in \cite{reichhardt13}. Current inversion was also reported in \cite{mcdermott16} for particles moving in periodic substrates at high densities.

This manuscript is organized as follows: our model system is presented in Sec. II. The numerical results and discussion are presented in Sec. III. Our conclusions are given in Sec. IV.

\section{Model}
\begin{figure}
  \includegraphics[scale=0.43]{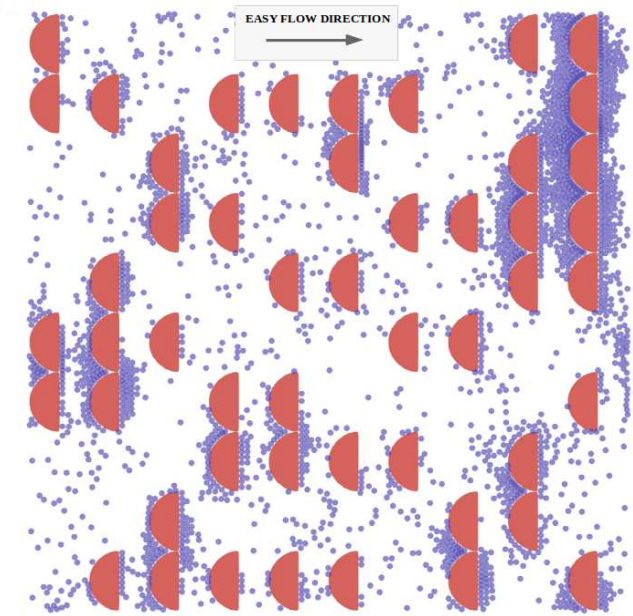}
  \caption{Snapshot of the system for a ratio of occupied area $\phi=0.3$ and fraction of removed obstacles $f=0.50$.The positive direction of mean particle current (easy flow direction) is also indicated.\label{fig1}}
\end{figure}

Our model consists of a two-dimensional (2D) system of $N$ soft active disks in a $L^2$ box, in which there are half-circles of diameter $D$ arranged in a square lattice of unit cell length $D$, initially with  $N_0=(L/D)^2$ obstacles. The obstacles are oriented in {\color{black} such} a way that the easy flow direction is the $+x$-direction. This means that neighboring obstacles touch each other along their diameters (y-axis), blocking the motion along x-axis, see Fig. \ref{fig1} . We introduce disorder by randomly removing a fraction $f$ of the obstacles.

The disks interact through a linear spring force law ${\bf F}_{ij}=\kappa(d_{ij}-r_{ij}){\hat{\bf r}}_{ij}$, for $r_{ij}<d_{ij}$ (otherwise, ${\bf F}_{ij}=0$), with $i\neq j=[1,N]$, $r_{ij}=|{\bf r}_i-{\bf r}_j|$ is the distance between particles, $d_{ij}=(d_i+d_j)/2$ is the mean diameter of a contact; for a disk-disk contact: $d_{ij}=d$; for a disk-obstacle contact: $d_{ij}=(d+D)/2$ for the curved side and $d_{ij}=d/2$ for the flat side. The arrange of the disks follow the usual set of active Langevin equations. For a given disk $i$ \cite{fily12}:
\begin{equation}
\label{eq1}
\frac{d{\bf r}_i}{dt}=\mu{\bf F}_i+{\bf v}_i+{\boldsymbol{\xi}}_i(t),~~~~~~
\frac{d\theta_i}{dt}=\eta_i(t),
\end{equation}
\noindent where $\mu$ is the motility, ${\bf F}_i=\sum\limits_{j}{\bf F}_{ij}$, ${\bf v}_i=v_o(\cos\theta_i{\bf\hat{i}}+\sin\theta_i{\bf\hat {j}})$ is the active velocity, $v_o$ is its magnitude and $\theta_i$ is its random direction; ${\boldsymbol{\xi}}_i(t)$ is a random thermal velocity, and $\eta_i$ is the random angular velocity. Both quantities are Gaussian white noises that follow $\left<{\boldsymbol{\xi}}_i(t)\right>=0$ and $\left<\xi_{i\alpha}(t)\xi_{j\beta}(t^\prime)\right>=(2\xi\Delta t)^{1/2}\delta_{ij}\delta_{\alpha\beta}\delta(t-t^\prime)$, $\alpha,\beta=x,y$ and $\left<\eta_i(t)\right>=0$ and $\left<\eta_i(t)\eta_j(t^\prime)\right>=(2\eta\Delta t)^{1/2}\delta_{ij}\delta(t-t^\prime)$; $\xi$ and $\eta$ are the noise intensities; since we consider the athermal model of \cite{fily12}, we set $\xi=0$. In all simulations we employ periodic boundary conditions (PBC)along $x$,$y$ directions. We integrate Eq. (\ref{eq1}) using a second order, stochastic Runge-Kutta algorithm \cite{honeycutt92}. The values of the model parameters are $d=1$ and $v_o=1$, which set length and time units, $D=10$, $\mu=1$, $L=100$, $\kappa=10$ (for a disk-obstacle contact $\kappa_{\mathrm{obs}}=1000$), $f=[0.05,1.00]$, with $\Delta f =$ $0.05$, and $\eta=0.001$.

We define here a mean net particle current (average velocity), as
\begin{equation}
\label{eq2}
{\bf J}=\frac{1}{N}\left<\sum\limits_{i=1}^{N}\frac{d{\bf r}_i}{dt}\right>,
\end{equation}
where the brackets denote averages over time and distinct realizations, as a function of $f$, and $\phi=N\pi/4L^2[1-(1-f)\pi/8]$, which is the area fraction defined as the ratio between the area of the disks by the available area {\color{black}(the difference between the total area and the remaining lattice area)}. In our simulations, we considered  $\phi=[0.1,0.9]$ and $ \Delta \phi = 0.1$.
\section{Results and discussion}
\begin{figure}[h]
 \includegraphics[scale=0.51]{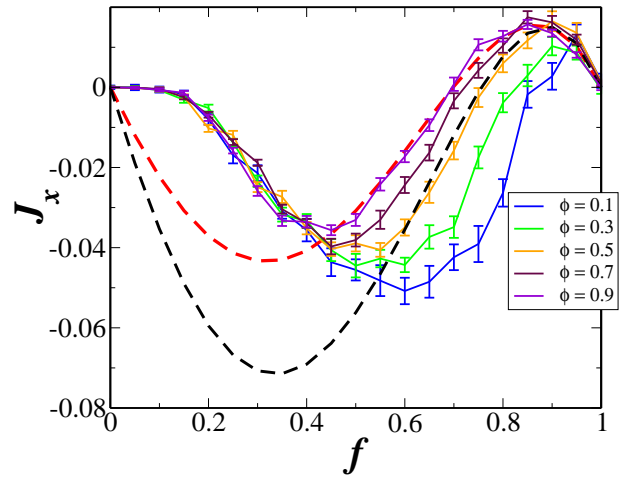}
  \caption{Mean net particle current $J_x$ as a function of $f$ for several area fractions $\phi$. The red and black dashed lines are obtained from Eq. (\ref{eq:theoJx}) for $\langle v\rangle=1.50\times10^{-4}$, $\langle c_+\rangle=2.00$, $\langle c_-\rangle=1.77$, and $\langle v\rangle=3.00\times10^{-4}$, $\langle c_+\rangle=1.35$, $\langle c_-\rangle=1.00$, respectively.\label{fig:meanJx}}
\end{figure}

{\color{black}In Fig. \ref{fig:meanJx} we show the $x$ component of the net particle current, i.e., $J_x$, as a function of $f$ for all $\phi$}. The mean perpendicular component, $J_y$, is zero in all cases and its results are omitted here. This occurs because the system is locally symmetric in $y$, although translational invariance is also broken along this direction. This indicates that to achieve a nonvanishing current, only {\em local symmetry} is sufficient.
%%%%%%%%%%%%%%%%FIGURA 2%%%%%%%%%%%%%%%%%%%%%%%
%%%%%%%%%%%%%%%%%%%%%%%%%%%%%%%%%%%%%%%%%%
As can be observed in Fig. \ref{fig:meanJx}, the mean current has two regimes: {\color{black}the first one, $J_x<0$ , for low and intermediate $f$}, indicating an {\em inversion} of the net particle current; the second regime is observed for high $f$, where the net current follows the easy flow direction, as reported before \cite{potiguar14}. The beginning of the negative regime has a vanishing current up to $f\approx0.10$. This is {\color{black} an} evidence of a minimum number of obstacles to be removed in order to have a non-vanishing current, since for $f=0$, there is no space for particles to move through the lattice. This minimum number, in this square lattice, equals the number of its columns. Hence, for our set up, we have $L/D$ columns, and the minimum fraction we need is $f_{min}=D/L=0.10$, which is close to the value we observed in our simulations. Note that for $f$ slightly larger than $f_{min}$ the current should not be very large since the majority of defective lattices will not have a percolated path along $x$.
These two regimes, negative and positive (easy flow direction) currents, {\color{black}change continuously, and it occurs at a $\phi$-dependent} fraction, $f^*(\phi)$, which decreases with increasing $\phi$. Also, the extrema of $J_x$ in both regimes, the negative minimum $J_{min}<0$, and the positive maximum $J_{max}$, also depend on $\phi$: as the area fraction increases, $|J_{min}|$ decreases as well as the $f$ value in which it occurs. For $J_{max}$, it increases with $\phi$, although not as strongly as the decrease of $|J_{min}|$, and its $f$ value decreases with $\phi$, just as its negative counterpart.

To explain the first regime ($J_x<0$), we go back to Fig. \ref{fig1}. We see that, where two obstacles touch, they form a region where there are, usually, some particles: we {\color{black}call} these regions as traps. Such accumulation was already seen in studies of the force between two bodies in a bath of active matter \cite{Ni2015,leite16}. Moreover, only particles drifting in the $+x$ direction are able to reach such traps by {\color{black}hitting} the obstacles, then sliding along their surfaces, a behavior largely reported for active matter \cite{drocco12-02,galajda07,potiguar14,bechinger16,volpe11,kaiser12,takatori14}. On the other side of the obstacles, there is a similar situation, the particles reach the flat {\color{black}sides} and slide along them, i.e., along the $\pm y$-directions. In this case, only particles traveling in the $-x$ direction will reach the flat sides of the obstacles. Therefore, we see that only those particles that are not trapped in any of these two structures will contribute to $J_x$. Therefore, we argue that there are more particles trapped in between the obstacles than the number of them stuck to the flat sides. This imbalance leads to more particles moving towards $-x$, yielding a negative net current.

In view of this idea, we obtained an expression relating $J_x$ to $f$ based on this argument (see details in Appendix). From Eq. (\ref{eq2}), we can write the mean net $x$-current as $J_x=\langle J_+\rangle-\langle J_-\rangle$, where $J_{\pm}$ are the currents in the $\pm x$-directions. We assume that such currents are given by the following expression
\begin{equation}
\label{eq:jxNV}
J_x=\langle n_+\rangle\langle v_+\rangle-\langle n_-\rangle\langle v_-\rangle,
\end{equation}
where $n_{\pm}$ are the number of particles that move in the $\pm x$ directions, and $v_{\pm}$ their respective velocities. Eq. (\ref{eq:jxNV}) takes into account that these two quantities are statistically independent. We further assume that both positive and negative mean velocities have the same value. {\color{black} We also assume reasonable dependencies on $f$ for $\langle v_{\pm}\rangle$, and we estimate the number of particles in each trap to the layers of particles that form around each obstacle (as seen in Fig. \ref{fig1}), and we end up with}:
\begin{equation}
\label{eq:theoJx}
J_x=v'\left(\frac{L}{D}\right)^2(1-f)f\left[\pi\frac{D+d}{2d}(f^2-1)\langle c_+\rangle+\frac{D}{d}\langle c_-\rangle\right],
\end{equation}
where $v'$ is the $x$-component of the self-propelling velocity, $\langle c_+\rangle$ is the mean number of layers of particles around the curved side, and $\langle c_-\rangle$ is the mean number of layers of particles around the flat side of each obstacle. In this calculation, we estimate the number of {\color{black}particles} per layer on the curved side to the value of the first layer of particles.
%In fig. \ref{fig:theoJx}, we show a few plots of this equation for a few values of $\left<c_{\pm}\right>$.

In Fig. \ref{fig:meanJx}, we also plot two examples obtained from Eq. \ref{eq:theoJx} that qualitatively reproduce the results above $J_{min}$ of the $\phi=0.90$ (red dashed) and $\phi = 0.50$ (black dashed) curves. The parameters we use to plot the theoretical results are, for the red curve,  $\langle v\rangle=1.50\times10^{-4}$, $\langle c_+\rangle=2.00$, and $\langle c_-\rangle=1.77$. For the black curve, we have $\langle v\rangle=3.00\times10^{-4}$, $\langle c_+\rangle=1.35$, $\langle c_-\rangle=1.00$. Notice that, in both cases $\langle c_+\rangle>\langle c_-\rangle$, which means that more particles are trapped in the spaces between the obstacles as compared to those stuck on the flat sides, which is consistent with our original argument. Also, we had to use a smaller mean velocity in order to reproduce the curve for larger $\phi$, which is also consistent with the fact that there are more particles in the system, and this will, eventually, {\color{black}diminish} the space available for motion, which might lead to a reduction of such velocity.

Eq. (\ref{eq:theoJx}) fails to reproduce the numerical results of $J_x$ below $J_{min}$. We argue that this is due to the breakdown of our first assumption, the independence between $\langle n_{\pm}\rangle$ and $\langle v_{\pm}\rangle$. We may understand this breakdown as the consequence of the small free space available for motion at such low $f$: there is a larger chance of occurrence of large obstacle clusters (see upper right of Fig. \ref{fig1} for an example of a 5-obstacle cluster); given that, most of the particles will be around an obstacle, and not able to freely move, and this will render our assumption incorrect. This dependence between $\langle n_{\pm}\rangle$ and $\langle v_{\pm}\rangle$ also explains the relation between $J_x$ and $\phi$ at low $f$, see below.

Along the same reasoning, we see that the $J_x>0$ range can be seen as the consequence of the more frequent occurrence of small cluster aggregates and isolated obstacles. We have checked this by generating several distinct disordered lattices for a given $f$ and measuring the proportion of cluster aggregates.

In each of such structures, there are less traps to hinder {\color{black} the} motion along the $+x$-direction, as discussed above, but there is still the flat side to block the path of particles moving along the $-x$-direction. Also, for an isolated obstacle, since particles slide along its surfaces, there is no {\color{black}restriction} for motion along $+x$-direction.
%%%%%%%%%%%%%%%%FIGURE 3%%%%%%%%%%%%%%%%%%%%%%%%%%%
\begin{figure}[h]
%\includegraphics[scale=0.8]{/home/fabricio/Alunos/Borba/jx_vx_Phi_1.eps}
%  \centering
    \includegraphics[scale=0.39]{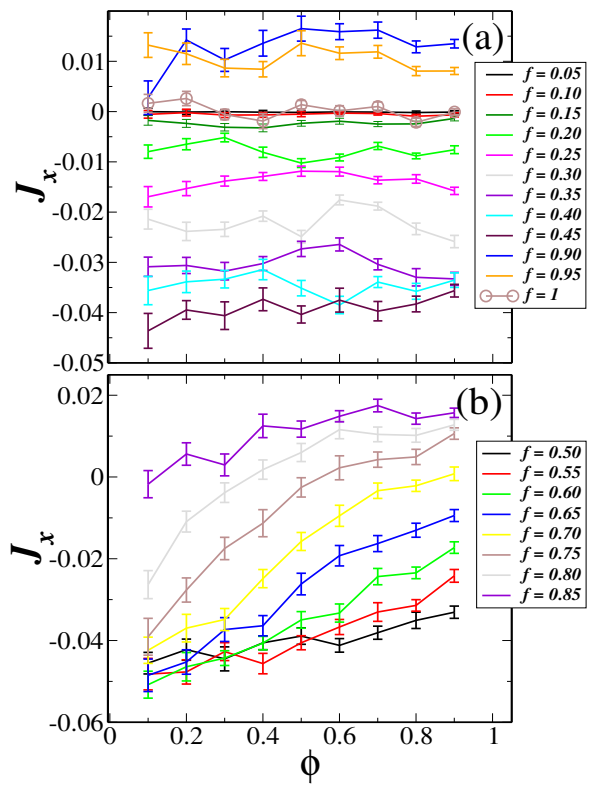}\\
  \caption{ The mean net particle current $J_x$ as a function of the area fraction $\phi$, for different values of $f$ in the two regimes: (a) $J_x$ independent of $\phi$, and (b)$J_x$ monotonically increasing with $\phi$.\label{fig:phiJx}}
\end{figure}
%%%%%%%%%%%%%%%%%%%%%%%%%%%%%%%%%%%%%%%%%%%%%%%%% 
Therefore, at this high $f$ regime, there are more particles traveling in the easy flow direction, yielding a positive current (see Appendix for other plots of Eq. (\ref{eq:theoJx})). 

{\color{black}In Fig. \ref{fig:phiJx}, we show the dependence of $J_x$ on $\phi$}. We identify two regimes for such dependence: one that the current is independent on $\phi$, Fig. \ref{fig:phiJx}(a), and the other in which $J_x$ increases with increasing $\phi$, Fig. \ref{fig:phiJx}(b). To understand these results, we see that the effect of adding more particles to this system is to increase the number of particles that contribute to $J_x$, as well as decreasing the free space for motion, since some of the {\color{black}particles} will get stuck in the traps in both sides of the obstacles. For low $f$, the free space is very limited, as stated above. Hence, a large number of particles that contribute to the current will be compensated by a smaller mean velocity, and $J_x$ will not depend on $\phi$. Note, again, that this explanation is, essentially, the breakdown of the assumption of the independence between the number of particles and their velocities. The independence of $f$ on $\phi$ is also seen at large $\phi$ and intermediate $f$, or for $f\to1$. In these cases, the previous explanation does not hold. But, for such a strong disorder, there will be a majority of isolated obstacles. Since a non-vanishing current is determined by the interaction with the curved side \cite{potiguar14}, it is reasonable to assume that such plateaus in the curves for Fig. \ref{fig:phiJx}(b), are due to a saturation of the isolated obstacles in their capacity to direct particle motion. If we had larger obstacles, we argue that such plateaus would occur at higher $\phi$. For larger $f$, where the {\color{black}aggregation is} smaller, and there is a growing number of isolated obstacles, as already indicated above, adding more particles will not sensibly decrease the free space, and the particle mean velocity will be unaffected by a larger $\phi$. In this case, before the aforementioned saturation limit, the current will increase with the number of particles, i.e., it will increase with $\phi$.

\begin{figure}[h]
%\includegraphics[scale=0.8]{/home/fabricio/Alunos/Borba/jx_vx_Phi_1.eps}
%  \centering
    \includegraphics[scale=0.5]{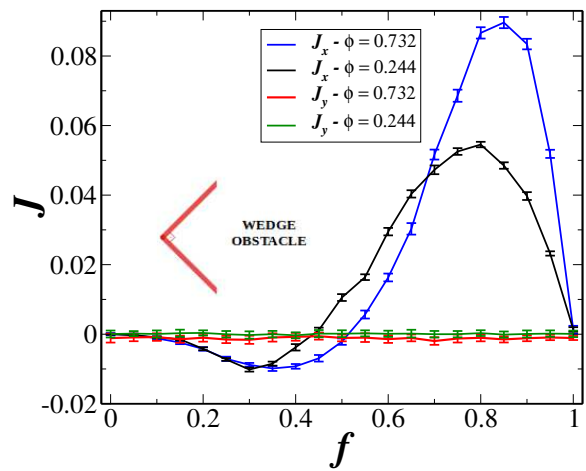}\\
  \caption{ The x and y components of the mean particle current, $J_x$ and $J_y$, as a function of $f$ for distinct values of the area fraction $\phi$. We consider wedges as obstacles. The aperture angle of the wedges is $90^\circ$. \label{fig:wedJx}}
\end{figure}

Finally, we varied the shape of the obstacle in order to investigate how distinct local symmetries affect the results: we used circles with $D=10$, and wedges whose sides are the ones of a square inscribed in the half-circles, and whose diagonal coincides with the diameter of the obstacles (this gives an angle of aperture of $90^\text{o}$). We also used half-circles of diameter $D=9$, so that we investigate whether the traps between the obstacles are really the responsible for the current inversion seen in Fig. \ref{fig:meanJx}. For the disordered lattice of circular objects, there is no net current in any case, even if we add an external field, the current will be only along this field. Again, local symmetry is the key ingredient to rectify the particle motion. For the disordered lattice of half-circles with $D=9$, thus allowing horizontal space between two adjacent obstacles, there is no spontaneous inversion of the current, we only observe $J_x>0$. This clearly validates our explanation that this phenomenon is due to the imbalance of particles traveling in both directions due to the traps provided by the obstacles. For the disordered lattice of wedges, there is the inversion, but the magnitude of the inverted (negative) current is smaller than that reported to the half-circles (see Fig. \ref{fig:wedJx}), while the positive currents are larger than those of Fig. \ref{fig:meanJx}. Besides, the minimum negative currents are rather insensitive to variations in density, while the maximum positive currents are larger for larger densities, a feature not seen in Fig. \ref{fig:wedJx}. The reason is that the concave side of the wedges trap more particles than the flat side of half-circles, reducing the current opposite to the easy flow direction. On the other hand, the convex sides of both wedge and half-circle obstacles present similar contribution for the particle current along the easy flow direction. Note that the opening angle of the wedges ($90^\circ$) favors the trapping \cite{kaiser12}. Still, in view of recent results of trapping of active matter in such {\color{black}funnel-shaped} obstacles \cite{kaiser12,kumar19}, there could be a richer phase diagram for this phenomenon, since the wedges have an additional degree of freedom, through their aperture angle, that can be explored.

%%%%%%%%%%%%%%%%FIGURE 4%%%%%%%%%%%%%%%%%%%%%%%%%%%
%%%%%%%%%%%%%%%%%%%%%%%%%%%%%%%%%%%%%%%%%%%%%%%%% 
\section{Conclusions}
We investigated the mean net current ${\bf J}$ (average velocity) of active particles interacting with a disordered square lattice of obstacles with the shape of half-circles. The diameter of the obstacles had the same size of the unit cell lattice, forming traps between them. The disorder was introduced by randomly removing a fraction of the obstacles. We observed that the net current along the easy flow direction at constant area fraction, for low and moderate disorder, presented a {\em spontaneous inversion}, i.e., particles traveled  in the negative easy flow direction, while at high enough disorder the net current was positive. We argued that such a phenomenon is the result of an imbalance of particles that get trapped in the spaces between the obstacles or get stuck in their flat side. We presented a theoretical calculation that reproduces the data reasonably well for $f$ above the value for which the current has a negative minimum, and it was based on the assumption of statistical independence between particle numbers and mean velocities. For a constant $f$, the dependence of $J_x$ on $\phi$ had two regimes: i) $J_x$ and $\phi$ are independent parameters; ii) the current grows with the area fraction. We argued that in the first regime, adding more particles, since the free space is {\color{black}smaller}, although {\color{black}that} would render more particles that contribute to the current, that would decrease their mean velocities in order that both effects compensate each other and the current does not change with $\phi$. In the second regime, this relation did not occur because there was more free space than {\color{black} in} the first regime, and $J_x$ grew with the number of particles. I.e., in the first case there was no statistical independence between the number of particles and their velocities, while in the second, there was such an independence. The direct (easy flow directions) and inverse particle currents, obtained and characterized in more details in a lattice of half-circles, were also observed in a lattice of wedges. In this case, the inverted current was smaller than the direct current, which was opposite to the behavior of the particle current in the lattice of convex obstacles (half-circles), and this was a consequence of the larger capacity of the wedges to retain particles in their concave side. Our results also point out the possibility of studying such an inversion phenomenon for a lattice of wedges that, given their additional degree of freedom (degree of aperture), could lead to a richer phase diagram for $J_x$ with $f$ and $\phi$. 
Our results are relevant for controlling active matter, e.g. in micro-fluidic devices. As an extension of the present work, the sorting of particles in a binary system can be studied. It is interesting to investigate how particles, e.g., with distinct noises, respond to local and global asymmetries such as the ones introduced here. Notice that high noise particles have less chance to stick in surfaces when compared to low noise particles, being less affected by the local asymmetry and opening the possibility to separate the distinct types of {\color{black}particles}, which is an experimentally demanding issue.
\section*{Appendix: derivation of $J_x$ in disordered lattices of asymmetrical obstacles}\label{Appendix}
The total current (velocity) in a system composed of N particles can be written as:
\begin{equation}\tag{A1}
%\mathbf{J} = \braket{\sum_{i = 1}^N \mathbf{v}_i}.
\mathbf{J} = \left<\sum_{i = 1}^N \mathbf{v}_i\right>.
\end{equation}
For our system, we are only interested in the x-component of this quantity; therefore, we write it as $J_x=\langle J_+\rangle\langle J_-\rangle$, where $J_{\pm}$ are the currents along the $\pm x$-directions. We assume that each of these currents is the product of the number of particles that contribute to its value, i.e., that are trapped, and their velocities, and that both quantities are statistically independent. Therefore, we have:
\begin{equation}\tag{A2}
\label{eq2}
J_x = \langle n_+\rangle\langle v_+\rangle-\langle n_-\rangle\langle v_-\rangle.
\end{equation}
Next, we consider that the averages of $\pm v$ are equal, such that the direction of the net current is determined only by the difference between the number of particles that travel in opposite directions:
\begin{equation}\tag{A3}
J_x = \langle v\rangle(\langle n_+\rangle-\langle n_-\rangle).\label{eq3}
\end{equation}
Assuming that, on average, half of the particles travel in each direction, we can write each of these average numbers as the difference between $N/2$ and the number of particles that are hindered in their motion due to the obstacles. Therefore, we have:
\begin{equation}\tag{A4}
\label{n+-}
\langle n_\pm\rangle = \frac{N}{2} - \langle T_\pm\rangle\langle p_\pm\rangle\textrm{,}
\end{equation}
where $T_\pm$ is the number of traps into which a particle might fall and $p_\pm$ is the number of particles in each trap. The number of traps, since they are formed by the obstacles, is a random number because it depends on a particular realization of the disordered obstacle lattice. 
%From now on, we assume that these two quantities are independent, and we end up with:
%\begin{equation}
%\Braket{n_\pm} = \frac{N}{2} - \Braket{T_\pm}\Braket{p_\pm}\textrm{.}\label{n+-}
%\end{equation}
For the negative direction, where the particles face the flat side of the obstacles, a trap is simply such section of a half-circle. Hence, for a given fraction of removed obstacles $f$, and since in the beginning we have $(L/D)^2$ obstacles in the lattice, where $D$ is the obstacle diameter, we obtain the following expression for the number of remaining traps in the lattice:
\begin{equation}\tag{A5}
\langle T_- \rangle = \left(\frac{L}{D}\right)^2(1-f).\label{t-} 
\end{equation}
For the positive direction, the traps are the spaces between two adjacent obstacles. For each removed obstacle, a trap will be removed as well, unless {\color{black}the obstacle is isolated or placed in between two other obstacles}. In the first case, no trap will be removed, while in the second, two traps will be removed. To take into account all these considerations, we simply assume that each removed obstacle corresponds to one removed trap. Therefore, the number of remaining traps for a given $f$ is the number of remaining obstacles. On the other hand, we know from \cite{potiguar14} that isolated obstacles do not trap any particle on the curved side. In fact, particles sliding on the curved side of the obstacle have their motion rectified along the $+x$-direction.

Based on our simulation results, we obtained the probability of occurrence of isolated obstacles as a function of the fraction of removed obstacles $f$. 

%%%%%FIGURE%%%%%%%%%%%%%%%%
\begin{figure}[h!]
\centering
\includegraphics[height=5cm]{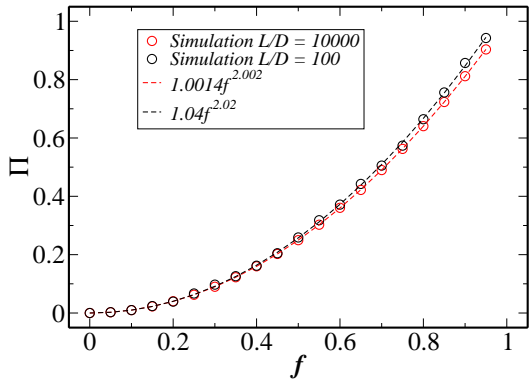}
  \caption{Ocurrence probability of isolated obstacles (diameter D) in the disordered lattice as a function of $f$ for two sizes L of the simulation box. The circles represent the simulation data and the dashed lines represent the fitting functions.}\label{isol}
\end{figure}
\begin{figure}[h!]
\centering
  \includegraphics[height=5.5cm]{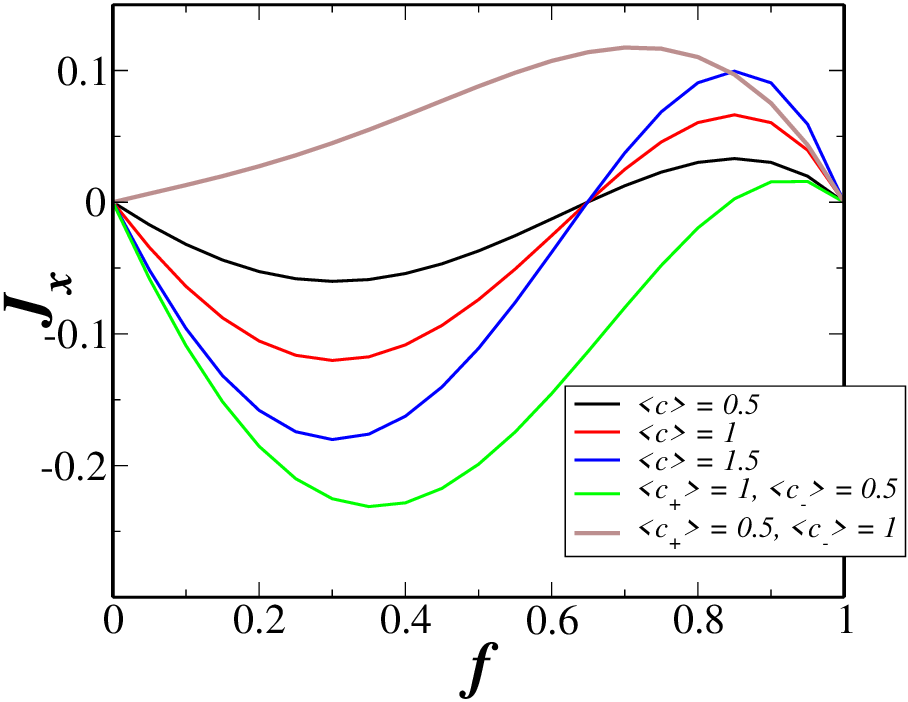}
  \caption{The mean net particle current $J_x$ as a function of $f$ for $\langle v\rangle=1.0\times10^{-4}$, and distinct values of the model parameters $\left<c_+\right>$, $\left<c_-\right>$, and $\langle c \rangle = \langle c_+ \rangle = \langle c_- \rangle$.\label{fgr:jx}}
\end{figure}

%%%%%%%%%%%%%%%%%%%%%%%%%%%%%%%%%%%%%%%%%
This is shown in Fig. \ref{isol}, where we observe that the probability of occurrence of isolated obstacles depends {\color{black} on} $f^2$. Therefore, the mean number of remaining isolated obstacles as a function of $f$ is $(L/D)^2(1-f)f^2$, and we define $\langle T_+ \rangle$ as the difference between the total number of remained obstacles and the number of isolated ones:
\begin{equation}\tag{A6}
\langle T_+ \rangle = \left(\frac{L}{D}\right)^2(1-f)(1-f^2).\label{t+}
\end{equation}
Combining Eqs. (\ref{n+-}), (\ref{t-}) and (\ref{t+}) in Eq. (\ref{eq3}), we have the following expression for the mean net particle current:
\begin{equation}\tag{A7}
J_x = \langle v \rangle\left(\frac{L}{D}\right)^2(1-f)\left[\left(f^2-1\right)\langle p_+ \rangle + \langle p_- \rangle\right].
\label{j1}
\end{equation}
We still need to formulate the dependence of $p_\pm$ as a function of $f$ and the packing fraction $\phi$ of the system. In order to estimate the number of particles per trap, we define the average number of layers of particles around the curved and flat sides of an obstacle, as $\langle c_+ \rangle$ and $\langle c_- \rangle$, respectively. Thus,
\begin{equation}\tag{A8}
\langle p_+ \rangle = \pi \frac{D+d}{2d}\langle c_+ \rangle,\quad \langle p_- \rangle = \frac{D}{d}\langle c_- \rangle,
\end{equation}
where both coefficients on these expressions are the number of particles per layer in each side of an obstacle; for the curved side, we approximate the number of particles per layer, which depends on the layer number, as the number of particles in the first layer. For a low and moderate densities, such approximation is valid. Another assumption is related to the fact that the velocity should depend on $f$, since if $f=0$ all paths are closed and the mean particle current vanishes; Therefore, we assume, for the sake of simplicity, that $<v> = v'f$, where $v'$ is the $x$-component of the self-propelling velocity, which it may also {\color{black}depends} on $\phi$ ( we ignore such a  dependency in this calculation).
\begin{equation}\tag{A9}
\hspace{-5.pt}J_x = v'\left(\frac{L}{D}\right)^2(1-f)f\left\lbrace\left[f^2-1\right] \\
\left(\pi\frac{D+d}{2d}\langle c_+ \rangle\right) + \frac{D}{d}\langle c_- \rangle\right\rbrace . \label{jfinal}
\end{equation}
In Fig. \ref{fgr:jx} we illustrate Eq. \ref{jfinal} as a function of $f$ for the condition $\langle c \rangle = \langle c_+ \rangle = \langle c_- \rangle$.

%%%%%%%%%%%%%%%%%FIGURE%%%%%%%%%%%%%%%%%%%%%%%%%%%%%
%%%%%%%%%%%%%%%%%%%%%%%%%%%%%%%%%%%%%%%%%%%%%%%%%%%%%
%%%%%%%%%%%%%%%%%%%%%FIGURE%%%%%%%%%%%%%%%%%%%%%%%%%%%%%%%%%%%%%
\begin{figure}[h!]
\centering
  \includegraphics[height=5.5cm]{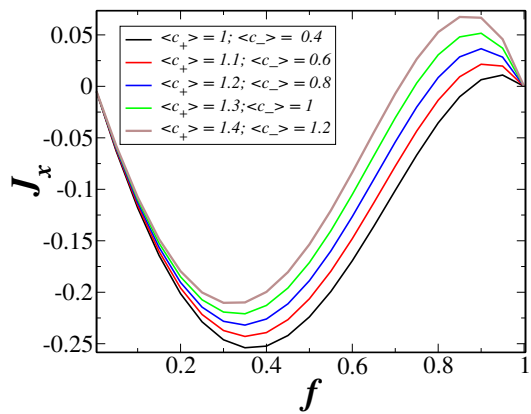}
  \caption{The mean net particle current $J_x$ as a function of $f$ for $\langle v\rangle=1.0\times10^{-4}$, and distinct values of the model parameters $\left<c_+\right>$, $\left<c_-\right>$.\label{jx2}}
\end{figure}
%%%%%%%%%%%%%%%%%%%%%%%%%%%%%%%%%%%%%%%%%%%%%%%%%%%%%%%%%%%%%%%
We expect $\langle c_\pm \rangle$ to be dependent on $\phi$.
Comparing with our numerical results, we observe that the values of $\langle c_\pm \rangle$ which better reproduce Fig. \ref{fig:meanJx} are those cases with $\langle c_+ \rangle > \langle c_- \rangle$, which is expected since we assumed that the mean velocities along each direction are equal. In Fig. \ref{jx2}, we show some cases according to such a condition. 

\section*{Acknowledgements}
The authors acknowledge financial support from CAPES, CNPq, FAPESPA, and FUNCAP (brazilian agencies); they also thank A. A. Moreira and J. S. Andrade Jr. for discussions and reading of the manuscript.
 

\begin{thebibliography}{99}

%\textcolor{blue}
\bibitem{toner05} J. Toner, Y. Tu and S. Ramaswamy, {\em Ann. Phys.} $\mathbf{318}$, 170 (2005).
\bibitem{ramaswamy10} S. Ramaswamy, {\em Ann. Rev. Cond. Matt. Phys.} {\bf 1}, 323 (2010).
\bibitem{vicsek12} T. Vicsek and A. Zafeiris, {\em Phys. Rep.} {\bf 517}, 71 (2012).
%\bibitem{marchetti12} M. C. Marchetti, J.-F. Joanny, S. Ramaswamy, T. B. Liverpool, J. Prost, M. Rao and R. Aditi Simha, cond.mat-soft/1207.2929 (2012).
\bibitem{marchetti12} M. C. Marchetti, J. F. Joanny, S. Ramaswamy, T. B. Liverpool, J. Prost, M. Rao and R. Aditi Simha, {\em Rev. Mod. Phys.} {\bf 85}, 1143 (2013).
\bibitem{bechinger16} C. Bechinger, R. Di Leonardo, H. Loewen, C. Reichhardt, G. Volpe and G. Volpe, {\em Rev. Mod. Phys.} {\bf 88}, 045006 (2016).
\bibitem{vicsek95} T. Vicsek, A. Czir\'ok, E. Ben-Jacob, I. Cohen, and O. Shochet, {\em Phys. Rev. Lett.} {\bf 75}, 1226 (1995).
\bibitem{drocco12-02} J. A. Drocco, C. J. Olson-Reichhardt, and C. Reichhardt,{\em Phys. Rev. E} {\bf 85}, 056102 (2012).
\bibitem{fily12} Y. Fily, and M. C. Marchetti, {\em Phys. Rev. Lett.} {\bf 108}, 235702 (2012).
\bibitem{wan08} M. B. Wan, C. J. Olson Reichhardt, Z. Nussinov, and C. Reichhardt,{\em Phys. Rev. Lett.} {\bf 101}, 018102 (2008).
\bibitem{tailleur09} J. Tailleur and M. E. Cates, {\em Europhys. Lett.} {\bf 86}, 60002 (2009).
\bibitem{galajda07} P. Galajda, J. Keymer, P. Chaikin and R. Austin, {\em J. Bacter.} {\bf 189}, 8704 (2007).
\bibitem{ghosh13}P. K. Ghosh, V. R. Misko, F. Marchesoni, and F. Nori, {\em Phys. Rev. Lett.} {\bf 110}, 268301 (2013).
\bibitem{reichhardt13}C. Reichhardt, and C. J. O. Reichhardt, {\em Phys. Rev. E} {\bf 88}, 062310 (2013).
\bibitem{potiguar14} F. Q. Potiguar, G. A. Farias and W. P. Ferreira, {\em Phys. Rev. E} {\bf 90}, 012307 (2014).
\bibitem{chepizhko13-1} O. Chepizkho, E. G. Altmann and F. Peruani, {\em Phys. Rev. Lett.} {\bf 110}, 238101 (2013).
\bibitem{dolai2018phase} P. Dolai,A. Simha, and S. Mishra, {\em Soft matter} $\mathbf{14}$, 29 (2018).
\bibitem{chepizhko13-2} O. Chepizkho and F. Peruani, {\em Phys. Rev. Lett.} {\bf 111}, 160604 (2013).
\bibitem{reichhardt14} C. Reichhardt and C. J. O. Reichhardt, {\em Phys. Rev. E} {\bf 90}, 012701 (2014).
\bibitem{morin16} A. Morin, N. Desreumaux, J.-B. Caussin and D. Bartolo, {\em Nat. Phys.} {\bf 13}, 63 (2017).
\bibitem{wang17} J. Wang, D. Zhang, B. Xia and W. Yu, {\em Soft Matter} {\bf 13}, 758 (2017).
\bibitem{yllanes2017} D. Yllanes, M. Leoni, and M. C. Marchetti, {\em New Journal of Physics.} $\mathbf{19}$, 10 (2017).
\bibitem{mcdermott16} D. McDermott, C. Reichhardt and C. J. O. Reichhardt, {\em Soft Matter} {\bf 12}, 8606 (2016).
%\bibitem{baskaran08}A. Baskaran, and M. C. Marchetti, Enhanced diffusion and ordering of self-propelled rods, Phys. Rev. Lett. {\bf 101}, 268101 (2008).
\bibitem{honeycutt92} R. L. Honeycutt,{\em Phys. Rev. A} {\bf 45}, 600 (1992).
\bibitem{Ni2015} R. Ni, M. A. Cohen Stuart, and P. G. Bolhuis, {\em Phys. Rev. Lett.} {\bf 114}, 018302 (2015).
\bibitem{leite16} L. R. Leite, D. Lucena, F. Q. Potiguar and W. P. Ferreira, {\em Phys. Rev. E} {\bf 94}, 062602 (2016).
\bibitem{volpe11} G. Volpe, I. Buttinoni, D. Vogt, H.-J. Kummerer, and C. Bechinger, {\em Soft Matter} {\bf 7}, 8810 (2011).
\bibitem{kaiser12} A. Kaiser, H. H. Wensink and H. Loewen, {\em Phys. Rev. Lett.} {\bf 108}, 268307 (2012).
\bibitem{takatori14} S. C. Takatori, W. Yan and J. F. Brady, {\em Phys. Rev. Lett.} {\bf 113}, 028103 (2014).
\bibitem{kumar19} N. Kumar, R. K. Gupta, H. Soni, S. Ramaswamy and A. K. Sood, {\em Phys. Rev. E} {\bf 99}, 032605 (2019).

\end{thebibliography}
\end{document}